\documentclass{article}

\usepackage[final]{neurips_2021}

\usepackage[utf8]{inputenc}
\usepackage[T1]{fontenc}
\usepackage{hyperref}
\usepackage{url}
\usepackage{booktabs}
\usepackage{amsfonts}
\usepackage{nicefrac}
\usepackage{microtype}
\usepackage{xcolor}
\usepackage{graphicx}
\usepackage{float}

\title{Continued domain-specific pre-training of protein language models for pMHC-I binding prediction}

\author{%
  Sergio E. Mares\thanks{Center for Computational Biology, University of California, Berkeley, CA, USA} \\
  Ariel Espinoza Weinberger\thanks{Department of Electrical Engineering and Computer Sciences, University of California, Berkeley, CA, USA} \\
  Nilah M. Ioannidis\thanks{Center for Computational Biology and Department of Electrical Engineering and Computer Sciences, University of California, Berkeley, CA, USA; Chan Zuckerberg Biohub, San Francisco, CA, USA}
}

\begin{document}

\maketitle

\begin{abstract}
Predicting peptide--major histocompatibility complex I (pMHC-I) binding affinity remains challenging due to extreme allelic diversity ($\sim$30,000 HLA alleles), severe data scarcity for most alleles, and noisy experimental measurements. Current methods particularly struggle with underrepresented alleles and quantitative binding prediction. We test whether domain-specific continued pre-training of protein language models is beneficial for their application to pMHC-I binding affinity prediction. Starting from ESM Cambrian (300M parameters), we perform masked-language modeling (MLM)-based continued pre-training on HLA-associated peptides (epitopes), testing two input formats: epitope sequences alone versus epitopes concatenated with HLA heavy chain sequences. We then fine-tune for functional IC$_{50}$ binding affinity prediction using only high-quality quantitative data, avoiding mass spectrometry biases that are inherited by existing methods.

\textbf{Key Results:} After continued pre-training and fine-tuning, our resulting model (ESMCBA) achieves a median Spearman correlation of 0.62 for predicting binding affinity across 25 common HLA alleles, outperforming NetMHCpan (0.56), MHCflurry (0.49), and other state-of-the-art predictors. Continued pre-training provides consistent gains relative to ESM Cambrian models that are directly fine-tuned without the continued pre-training step, particularly for alleles with a moderate amount of available binding data (500--2000 peptides), improving correlations by $\sim$0.10 over models without continued pre-training.

\textbf{Limitations:} The benefits of continued pre-training drop significantly for data-scarce alleles ($<$500 peptides), where models without continued pre-training outperform continued pretraining models. In addition, the method requires substantial computational resources (300M parameters), and the fine-tuning step remains limited by the inherent noise in binding affinity measurements. Binding prediction shows variable performance across alleles, highlighting ongoing challenges for generalization over data-scarce alleles.

\textbf{Impact:} This work has important potential application to neoantigen vaccine prioritization and provides a framework for improving protein language model performance on specialized tasks through domain-specific continued pre-training.
\end{abstract}

\section{Introduction}

Protein language models (PLMs) trained on large protein corpora have become foundational tools for structure and function prediction \citep{rives2021, lin2023}. However, most applications of these models to downstream tasks involve a standard fine-tuning approach. In natural language processing, domain-specific continued pre-training—where models undergo additional unsupervised training on task-relevant data before supervised fine-tuning—often yields substantial performance gains \citep{gururangan2020}. Whether this strategy translates effectively to protein modeling remains largely unexplored.

We investigate this question using peptide–major histocompatibility complex class I (pMHC-I) binding affinity prediction. This task represents an important test case for several reasons. First, accurate pMHC-I binding prediction is critical for vaccine design and personalized immunotherapy \citep{vita2019}, making performance improvements directly clinically relevant. Second, the task suffers from extreme data scarcity and imbalance: while humans express approximately 30 000 different HLA class I alleles, quantitative binding data exist for fewer than 200 alleles, with most alleles having fewer than 1000 measured peptide binding affinities. Third, the task requires joint modeling of highly polymorphic HLA chains and diverse peptide sequences, creating a stringent benchmark for cross-sequence generalization.

Current pMHC-I binding predictors face three fundamental challenges. \textbf{Allelic diversity:} The extreme polymorphism of HLA genes creates a long-tail distribution where most alleles lack sufficient training data for robust supervised learning. \textbf{Experimental bias:} Mass spectrometry-based datasets systematically over-represent peptides with canonical anchor residues, creating training distributions skewed toward specific motifs while under-sampling weak binders \citep{bruno2023}. \textbf{Label heterogeneity:} Binding measurements come from diverse experimental protocols (competitive binding, mass spectrometry, fluorescence polarization) with varying quality and interpretation, complicating model training and evaluation.

\subsection{Hypothesis and Approach}

We hypothesize that domain-specific continued pre-training can improve protein language model representations of peptide sequences bound to MHC-I, improving downstream performance on binding affinity prediction. Specifically, we test whether additional masked-language modeling pre-training on HLA-associated peptides—before supervised fine-tuning—enables models to learn generalizable binding motifs across alleles.

Starting from the 300M-parameter ESM Cambrian model \citep{nijkamp2024, hayes2025}, we implement a two-stage training protocol:

\textbf{Stage 1 (Unsupervised):} Continued masked-language modeling pre-training on two domain-specific corpora: (i) epitope sequences alone and (ii) epitopes concatenated with their corresponding HLA heavy chains. 

\textbf{Stage 2 (Supervised):} Fine-tuning of the continued pre-training models for half-maximal inhibitory concentration  (IC$_{50}$) binding affinity prediction. To mitigate experimental bias, we train exclusively on high-quality functional antagonist assays, avoiding mass spectrometry data.

We evaluate our approach—termed ESMCBA (ESM Cambrian Binding Affinity)—on the hypothesis that continued pre-training should: (1) improve performance over baseline ESM models without additional pre-training, (2) enhance data efficiency for low-resource alleles, and (3) match or exceed current state-of-the-art predictors.

\subsection{Related Work}

Early pMHC-I binding predictors relied on position-weight matrices and linear models \citep{chen2019}. Modern neural approaches, including MHCflurry \citep{odonnell2020}, HLAthena \citep{sarkizova2020}, MHCnuggets \citep{shao2020}, NetMHCpan \citep{reynisson2020}, and HLApollo \citep{thrift2024}, have achieved substantial improvements. However, all of these methods train on the same types of experimental datasets and thus inherit the systematic biases present in mass spectrometry-derived training data \citep{bruno2023}.
Recent work has begun exploring protein language models for immunological applications. However, these efforts have primarily focused on standard feature-extraction approaches without investigating domain-specific continued pre-training \cite{thrift2024}. Our work fills this gap by systematically evaluating whether additional unsupervised learning on immunological sequences can improve downstream task performance.

In this work, we demonstrate that domain-specific continued pre-training significantly enhances pMHC-I binding prediction and outperforms existing methods. Additionally, we provide the first systematic analysis of how continued pre-training influences protein language models across varying conditions, showing that improvements are most pronounced for alleles with moderate data availability (500--2000 peptides). We introduce ESMCBA as a practical tool designed for neoantigen prioritization, effectively addressing critical limitations of existing predictors, particularly for underrepresented alleles. Finally, we establish a methodological framework for applying continued pre-training to specialized biological prediction tasks, providing guidance for future protein language model applications.

\section{Results}

\begin{figure}[ht]
    \centering

    \begin{minipage}[t]{0.35\textwidth}
        \centering
        \includegraphics[width=\textwidth,
                     trim=5 0 5 0, clip]{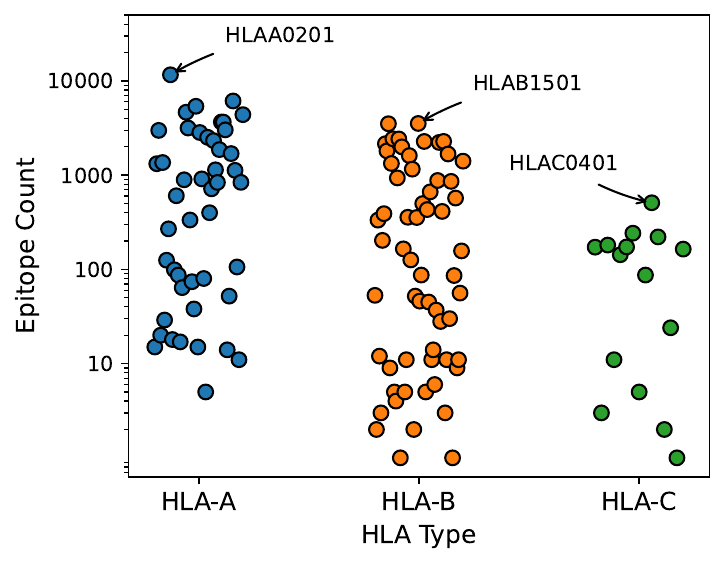}
    \end{minipage}%
    \hfill%
    \begin{minipage}[t]{0.6\textwidth}
        \centering
        \includegraphics[width=\textwidth]{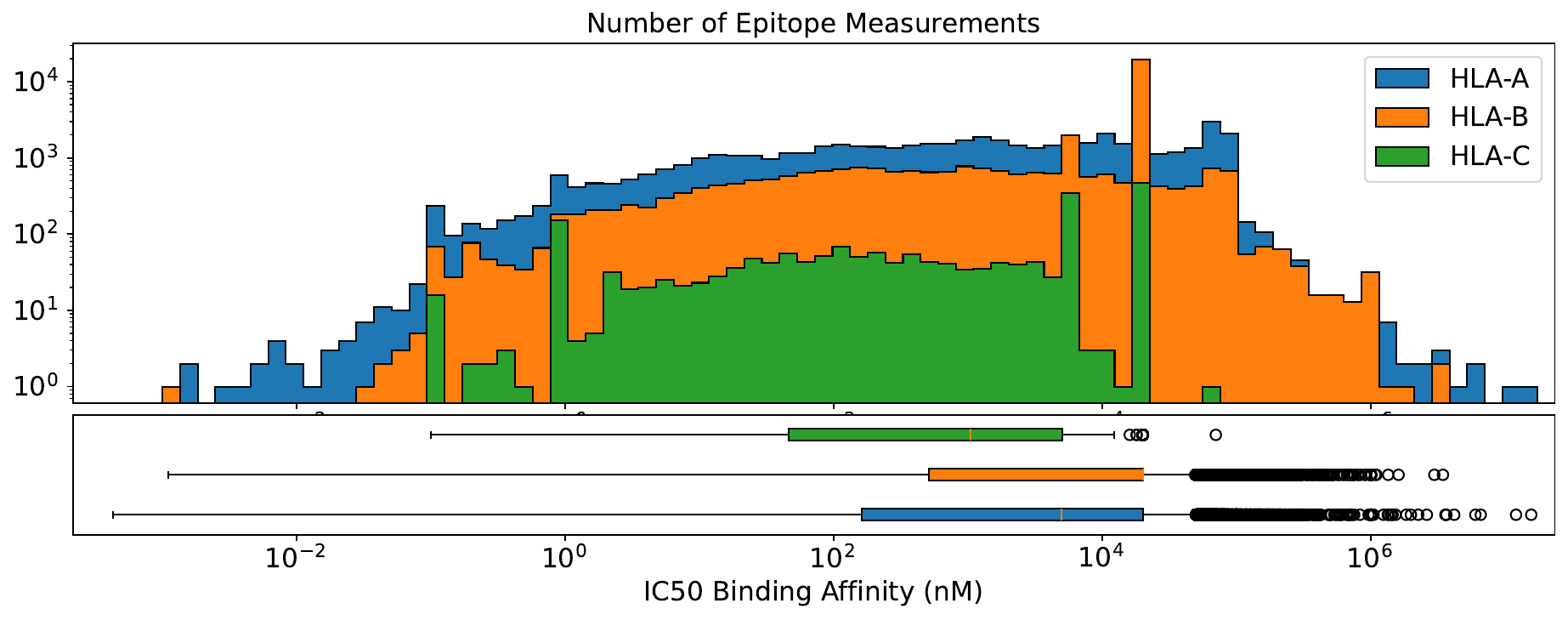}
    \end{minipage}
    \includegraphics[width=1\textwidth, trim=100 200 0 100, clip]{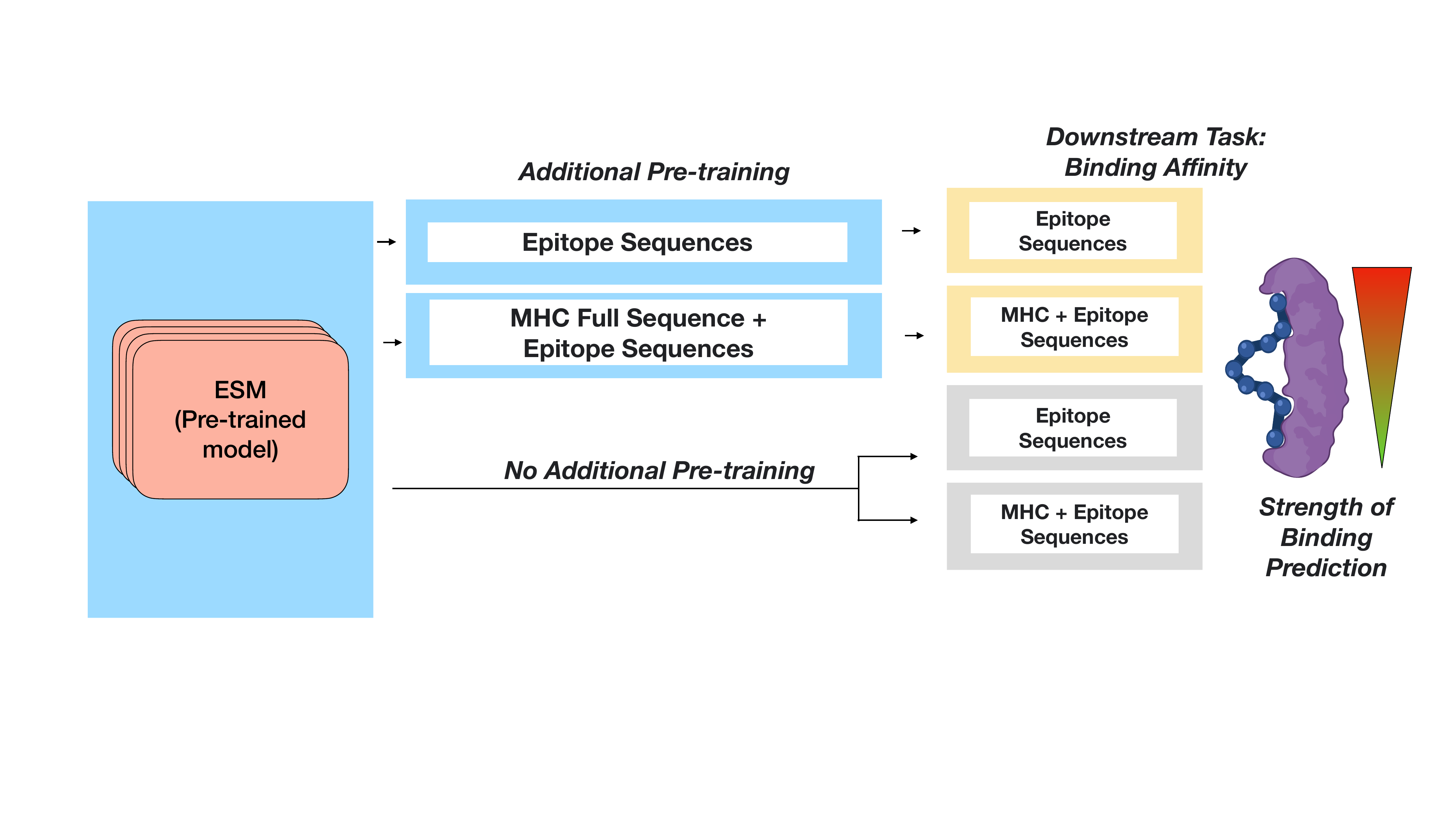}

    \caption{(A) Distribution of available pMHC-I training data across 121 HLA alleles from classes A, B, and C. (B) IC$_{50}$ binding affinity distribution (nM). (C) Two-stage training workflow: unsupervised continued pretraining on epitope sequences (E) or HLA+epitope concatenations (H+E), followed by supervised fine-tuning for binding affinity prediction. }
    \label{fig:distance}
\end{figure}

\subsection{IC$_{50}$ data is scarce across alleles}

We first extracted quantitative peptide–MHC binding affinity measurements from the Immune Epitope Database (IEDB) across available HLA-A, HLA-B, and HLA-C alleles, filtering entries to exclude sequences containing non-canonical residues. We show (Fig. \ref{fig:distance}a) that most alleles have fewer than 1000 associated peptide measurements, highlighting the data scarcity for many alleles. We also observe substantial variability and notable outliers in the measured IC$_{50}$ binding affinities (Fig. \ref{fig:distance}b); therefore, we log-transform the data to stabilize variance and improve downstream modeling performance.

\subsection{Continued pre-training leads to more accurate models}

We use an optional continued pre-training step followed by fine-tuning to develop ESMCBA for predicting peptide binding affinity across HLA alleles (Methods). Table~\ref{tab:mean_correlations} reports mean Spearman ($\rho$) and Pearson ($r$) correlations for five training–set sizes (number of peptides with binding affinity measurements available per allele) and reveals three clear trends.  
\textbf{(i) Very sparse data} ($<$500 peptides): the epitope-only model without continued pre-training but with full fine-tuning of all layers (\textsc{Non-PT E}) performs best ($\rho=0.33$, $r=0.32$), suggesting that when examples are limited the continued pre-training is not advantageous.  
\textbf{(ii) Moderate data} (500–2000 peptides): continued pre-training yields gains of roughly 0.10 in both $\rho$ and $r$—the \textsc{PT E} model benefits from unsupervised motif learning unavailable to its counterpart without continued pre-training.  
\textbf{(iii) Large data} ($>$2000 peptides): the HLA-concatenated model without continued pre-training (\textsc{Non-PT H}) matches or slightly surpasses \textsc{PT E}. Including the HLA sequence may shift the input length and amino acid sequence properties toward those encountered during the original ESM training, aligning the continued pre-training data with the original pre-training distribution.

\begin{figure}[htp]
  \centering
  \includegraphics[width=0.9\linewidth]{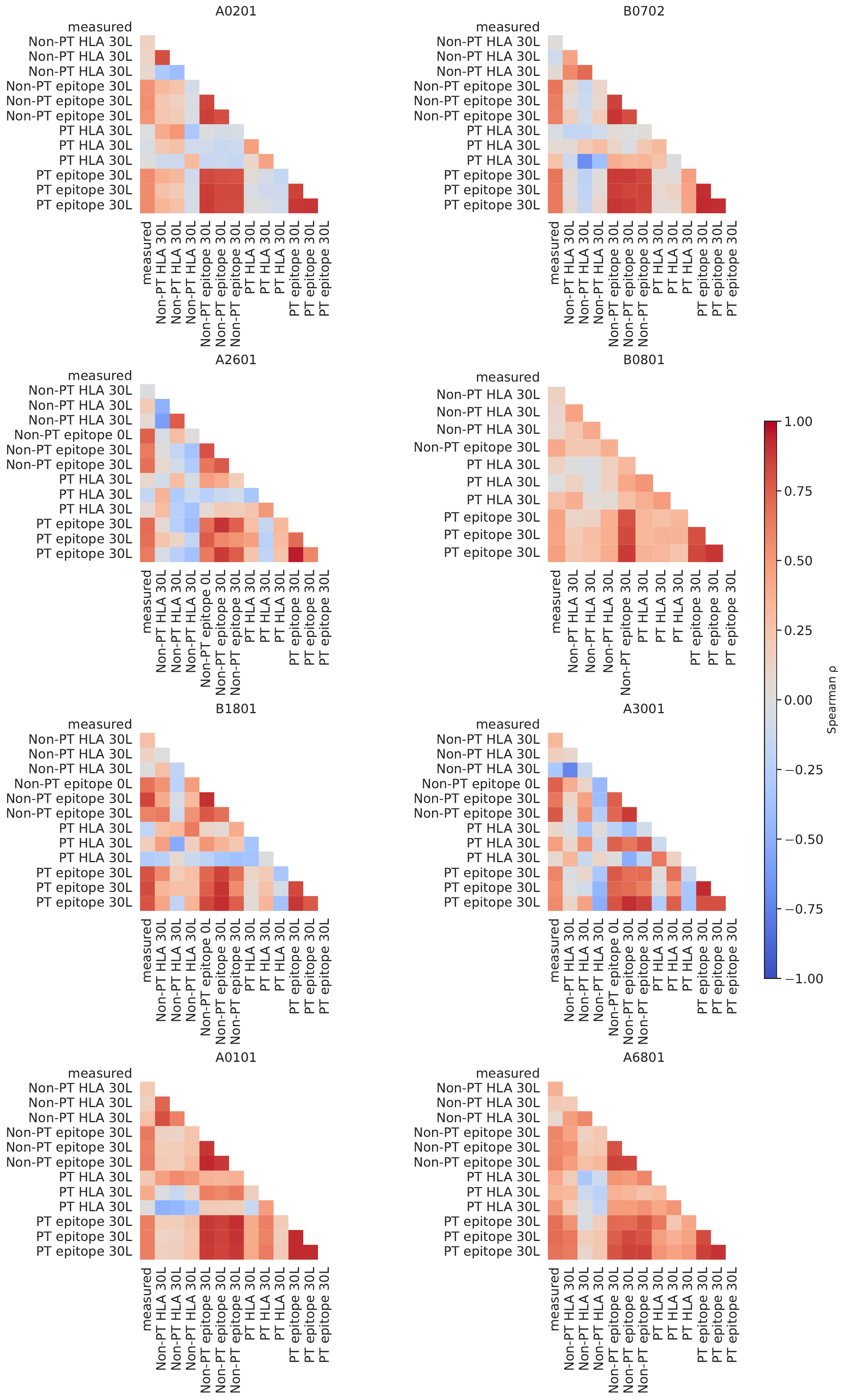}
\caption{Performance (Spearman correlation) matrices between the top model replicates (by spearman) across 9 representative HLA alleles evaluated on the test set. The measured label corresponds to the true binding affinity measurements. Each panel displays a correlation matrix for a specific allele, comparing predictions from different model variations.}

  \label{fig:matrix_correlations}
\end{figure}

We next compare ESMCBA predictions with measured IC$_{50}$ values for nine representative HLA alleles (Fig.~\ref{fig:matrix_correlations}). Four model configurations are shown—\textsc{Non-PT E 30L}, \textsc{Non-PT H 30L}, \textsc{PT E 30L}, and \textsc{PT H 30L}—with three replicates each. The pretrained epitope-only model (\textsc{PT E 30L}) aligns most closely with ground truth across alleles, achieving $\rho > 0.6$ for large-data alleles such as A*02:01 and B*07:02. Performance is more variable for under-represented alleles (e.g., A*30:01, B*08:01), underscoring the persistent challenge of accurate prediction in low-data regimes.

\begin{table}[ht]
  \centering
  \small
\caption{Mean Spearman $\rho$ and Pearson $r$ by model selection and training-set size.}
\label{tab:mean_correlations}
\begin{tabular}{lcccccccccc}
\toprule
 & \multicolumn{10}{c}{Size of Binding Affinity Training Data (Peptides)} \\
\midrule
 & \multicolumn{2}{c}{\textbf{$<$500}} & \multicolumn{2}{c}{\textbf{500--1000}} & \multicolumn{2}{c}{\textbf{1000--2000}} & \multicolumn{2}{c}{\textbf{2000--4000}} & \multicolumn{2}{c}{\textbf{$>$4000}} \\
\textbf{Model} & \boldmath$\rho$ & \boldmath$r$ & \boldmath$\rho$ & \boldmath$r$ & \boldmath$\rho$ & \boldmath$r$ & \boldmath$\rho$ & \boldmath$r$ & \boldmath$\rho$ & \boldmath$r$ \\
\midrule
\multicolumn{11}{l}{\textbf{\textsc{ESMCBA\textsubscript{E}} (E) Models}} \\
Non-PT E 0L  & 0.241 & 0.228 & 0.205 & 0.233 & 0.237 & 0.263 & 0.337 & 0.357 & 0.301 & 0.284 \\
Non-PT E 30L & \textbf{0.325} & \textbf{0.324} & 0.378 & 0.465 & 0.459 & 0.517 & 0.557 & 0.617 & 0.585 & 0.594 \\
PT E 0L      & -0.019 & 0.020 & 0.159 & 0.182 & 0.247 & 0.313 & 0.308 & 0.344 & 0.371 & 0.402 \\
PT E 30L     & 0.279 & 0.257 & \textbf{0.480} & \textbf{0.550} & \textbf{0.497} & \textbf{0.537} & \textbf{0.573} & \textbf{0.645} & \textbf{0.623} & \textbf{0.611} \\
\midrule
\multicolumn{11}{l}{\textbf{\textsc{ESMCBA\textsubscript{HLA+E}} (H) Models}} \\
Non-PT H 0L  & -0.064 & -0.041 & -0.037 & 0.034 & 0.002 & 0.039 & 0.052 & 0.096 & 0.167 & 0.187 \\
Non-PT H 30L & \textbf{0.177} & \textbf{0.195} & \textbf{0.433} & \textbf{0.500} & \textbf{0.530} & \textbf{0.568} & \textbf{0.569} & \textbf{0.634} & \textbf{0.637} & \textbf{0.624} \\
PT H 0L      & -0.067 & -0.031 & -0.157 & -0.118 & 0.025 & -0.007 & 0.108 & 0.080 & 0.150 & 0.149 \\
PT H 30L     & 0.155 & 0.115 & 0.267 & 0.347 & 0.443 & 0.502 & 0.534 & 0.597 & 0.608 & 0.597 \\
\bottomrule
\end{tabular}
\end{table}

\begin{figure}[htp]
  \centering
  \includegraphics[width=0.9\linewidth]{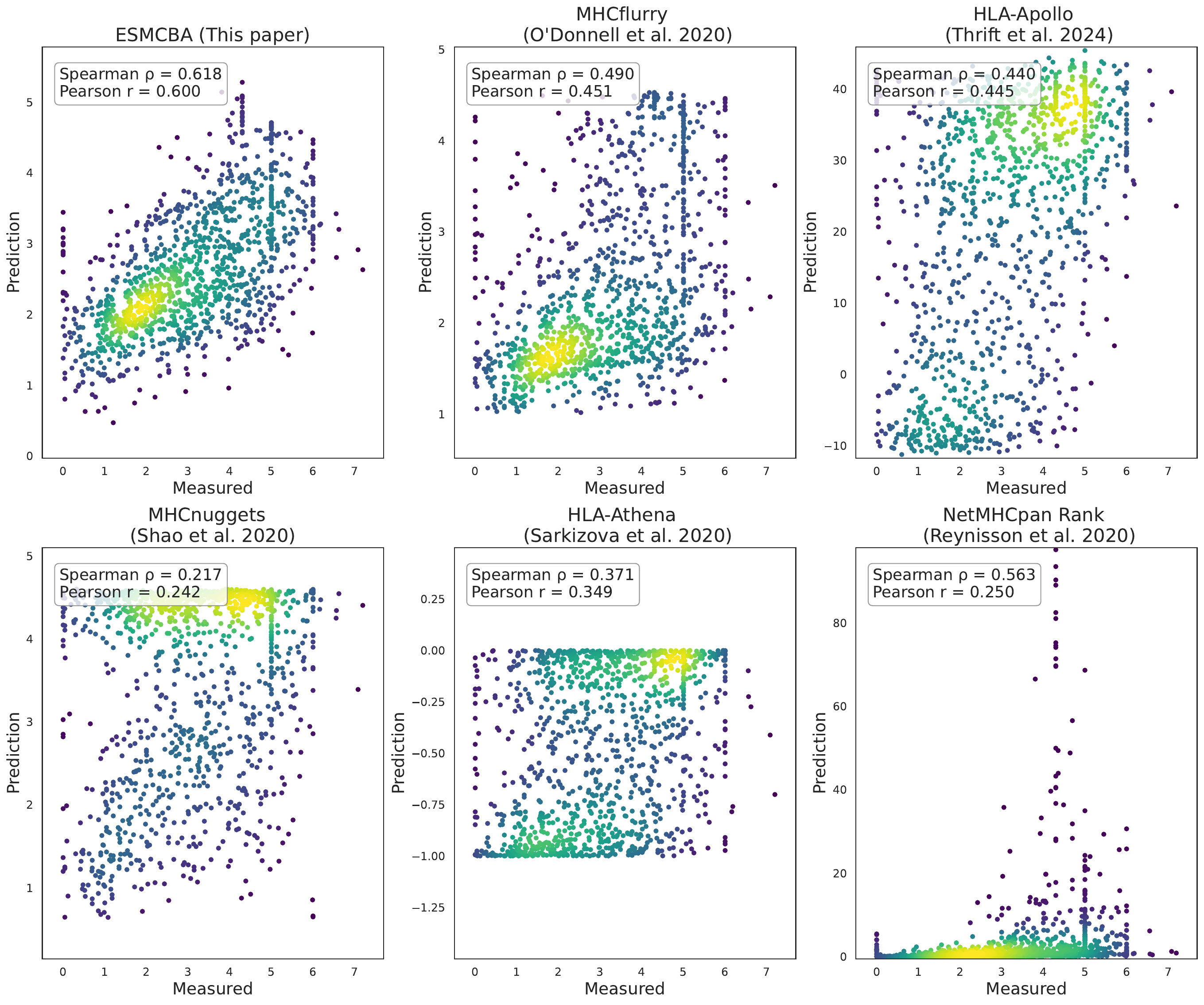}
  \caption{Predicted versus measured binding affinities for epitopes deposited in IEDB between 2020 and 2025 (held-out test set n=1,879 peptides).}
  \label{fig:ba_predictions}
\end{figure}

\subsection{Comparison to state-of-the-art models}
\label{subsec:results:ba_predictions}

To benchmark against state-of-the-art methods, we selected five widely adopted pMHC-I binding predictors: MHCflurry 2.0 \citet{odonnell2020}, NetMHCpan 4.1 \citet{reynisson2020}, HLAthena \cite{sarkizova2020}, HLApollo \cite{thrift2024}, and MHCnuggets \cite{shao2020}. We evaluated all models on the same held-out test set containing peptides deposited in IEDB after January 1, 2020. ESMCBA achieves a median Spearman correlation of $\rho = 0.62$ across 25 common HLA alleles, substantially outperforming all baselines on quantitative IC$_{50}$ binding affinity prediction (NetMHCpan: $\rho = 0.56$, MHCflurry: $\rho = 0.49$, HLApollo: $\rho = 0.44$, HLAthena: $\rho = 0.37$, MHCnuggets: $\rho = 0.22$). These gains indicate that domain-specific continued pre-training unlocks task-relevant sequence features, yielding consistently improved affinity predictions across alleles.

\subsection*{2.4 Allele-wise evaluation on qualitative assays}
\label{subsec:results:sensitivity}

To test how well each model generalizes beyond quantitative IC\textsubscript{50} labels, we assembled a held-out set of 18\,269 peptide–allele pairs that carry \emph{qualitative} annotations. To prevent any data leakage, we downloaded the original training sets used by ESMCBA and MHCFlurry and verified that none of the qualitative epitopes appeared in their training data.  These labels—\textit{Negative}, \textit{Positive-Low}, \textit{Positive-Intermediate}, \textit{Positive-High}, or \textit{Positive}—originate from diverse experimental protocols, including mass spectrometry, competitive-binding, and fluorescence-polarisation assays, and are therefore considerably noisier than the binding-affinity measurements used in Sections ~\ref{subsec:results:ba_predictions}.  For evaluation, we converted each ordered class boundary into a binary classification task and computed the per-allele AUROC before averaging across alleles.  

\begin{figure}[ht]
\centering
\includegraphics[width=1\linewidth]{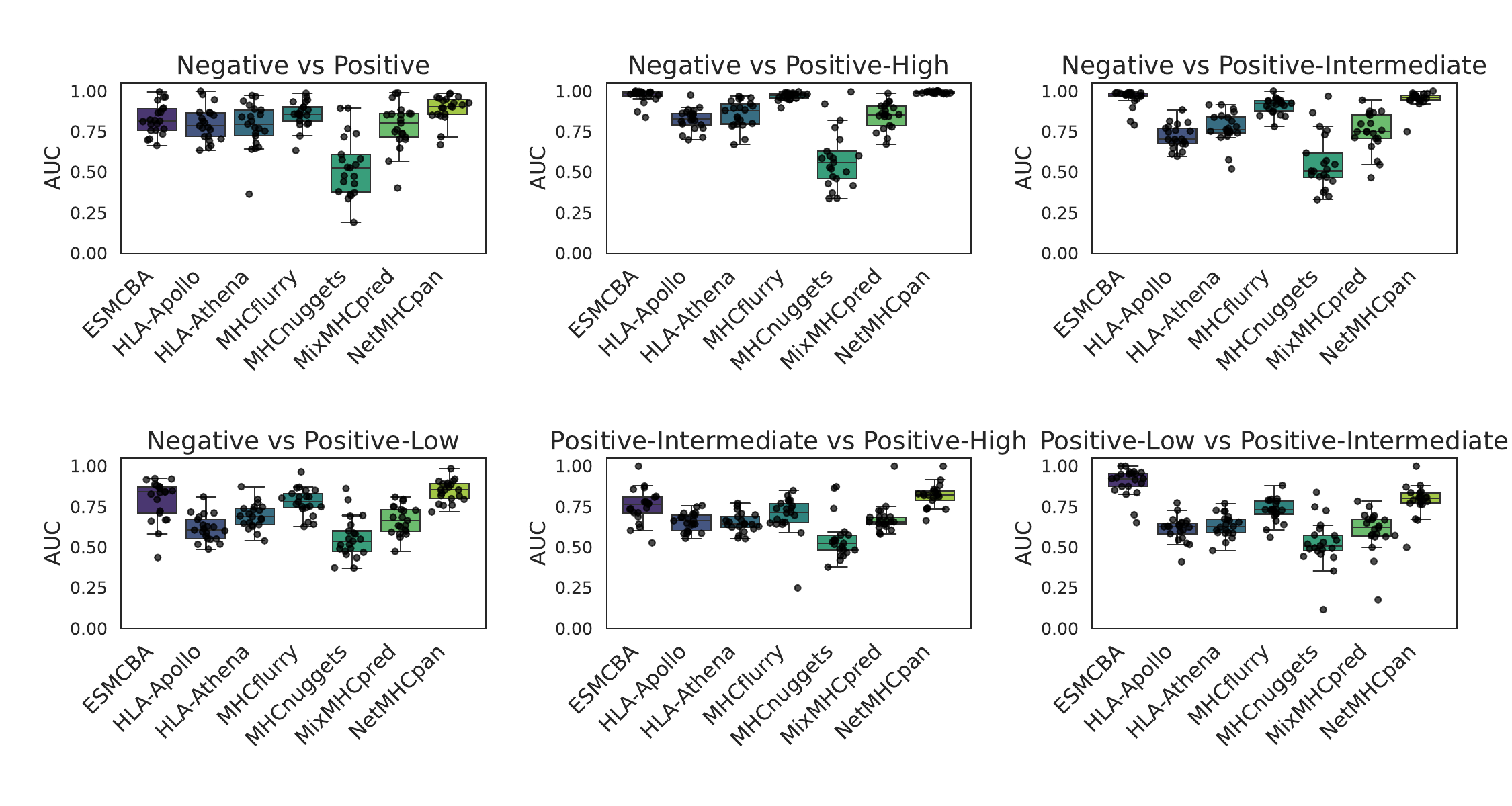}
\caption{\textbf{ROC-AUC performance across qualitative assay outcomes.} Evaluations based on 18,269 qualitative entries from IEDB, excluded from quantitative training sets.}
\label{fig:qualitative_auc}
\end{figure}

For the two most practically relevant screens—\textit{Negative vs Positive-Low} and \textit{Negative vs Positive-High}—ESMCBA achieves mean AUROC of 0.79 and 0.97, respectively, outperforming most models (Fig.~\ref{fig:qualitative_auc}), except for NetMHCpan (AUROC = 0.85). For the \textit{Positive-Low vs Positive-Intermediate} split, ESMCBA’s mean AUROC (0.95) is 0.11 higher than the next-best model, suggesting finer discrimination among weak binders. %This task is relevant as in a screen for prioritization for vaccines, discrimination to the strongest binder is key for immune presentation. 

Taken together, the qualitative benchmark confirms that continued pre-training confers measurable benefits even under substantial label noise. Our analyses highlight allele-specific data scarcity as a principal source of residual error, an issue future work may address with improved methods to model sequences that are out-of-distribution from previously tested peptides.

\section{Discussion}

\subsection{Improved workflow for pMHC binding prediction}

Our study demonstrates the benefits of extending domain-specific continued pre-training from natural language processing to protein modeling, specifically in the data-scarce landscape of immunology. Continued pre-training on domain-specific sequences improves predictive performance over traditional pMHC predictors, many of which are data-hungry or learn experimental biases such as those present in mass spectrometry data.

\subsection{Mechanisms behind the success of continued pre-training}

We propose two complementary mechanisms underlying these improvements. Firstly, pre-training on HLA-associated peptides may adjust the model's biochemical priors, better capturing residue preferences and interactions within binding peptides. Secondly, concatenating peptides with their corresponding HLA chains may facilitate learning of allele-specific binding contexts.

\subsection{Limitations and future work}

Data bias is a recurrent problem in immunopeptidomics and modeling of pMHC interactions. IEDB serves as a cornerstone for the development of these models, yet careful considerations and practices involving noise and false positives need to be taken into account for model improvements. Our current study does not explore variations in model scale or structural supervision, nor does it thoroughly address the noisy nature of qualitative labels. Future research could leverage this work and expand the framework to MHC class II and TCR-pMHC complexes, testing whether the continued pre-training approach is scalable and effective across broader immunological scenarios.

\subsection{Broader implications}

Our results underscore the value of continued pre-training when applying large PLMs for biochemical prediction tasks. Modest, targeted domain-specific pre-training can result in substantial improvements, providing a practical approach for developing predictive tools essential for personalized immunotherapies and neo-antigen discovery.

\section{Conclusion}

We present ESMCBA as a novel allele-aware extension of the ESM protein language models, enhanced by domain-specific continued pre-training specifically on peptide–MHC sequence data. Our approach incorporates only high-quality quantitative IC$_{50}$ measurements. By fine-tuning the task-relevant transformer layers, we significantly improve data efficiency and predictive accuracy. ESMCBA achieves a median Spearman correlation of 0.62 for binding affinity prediction across 25 common alleles, outperforming existing state-of-the-art predictors. The model also robustly generalizes to noisy qualitative labels, demonstrating resilience to experimental variability. Our results have important practical implications for accelerating neoantigen vaccine design cycles and facilitating large-scale screening for underrepresented alleles.

\section{Methods}

\subsection{Data curation}
\label{subsec:data_prep}
We applied the following pipeline to curate quantitative binding data from IEDB:
\begin{enumerate}
  \item Download raw IEDB entries (accessed on 16-01-2025) and filter peptides to lengths between 8 and 15 amino acids.
  \item Remove all entries containing non-canonical residues.
  \item Apply a log$_{10}$ transform to IC$_{50}$ values to stabilize variance.
  \item Perform a temporal split: all peptides submitted before January~1, 2020, were used for training; peptides submitted on or after January~1, 2020, were held out as a test set.
  \item Divide functional antagonist measurement log$_{10}$IC$_{50}$ values and subsample using a Gaussian kernel centered at $10^{3}$\,nM (the approximate mean affinity across alleles), which reduces class imbalance between high-affinity and low-affinity peptides. 
\end{enumerate}

\subsection{Unsupervised Continuation pre-training}
\label{subsec:mlm_pretrain}

ESM Cambrian model weights were downloaded from \url{https://github.com/evolutionaryscale/esm}. Sequences were tokenised with the 33-character ESM vocabulary and truncated or zero-padded to a maximum length of 1,024 tokens.
To adapt representations to allele-specific context, we continue pre-training with a masked-language-model (MLM) objective on peptide sequences or HLA-concatenated peptides. A linear head predicts the original amino acid at 15\% of randomly selected peptide positions, while HLA residues remain visible. From the IEDB, we used positive binders as described in the qualitative labels. Training sequences were split 80:10:10 into train, validation, and evaluation sets. Peptide and HLA tokens share the same vocabulary. We introduced data augmentation by duplicating the number of sequences in our training data for this step. 
The unsupervised continuation was trained for 10 epochs on a single RTX~2080 Ti GPU.

\subsection{Supervised binding–affinity fine-tuning}
\label{subsec:ba_finetune}

We attach a single-unit linear head to the 300 M-parameter \textit{ESM-Cambrian} backbone and unfreeze the 30 transformer blocks plus the final layer norm. The head receives the mean-pooled token embeddings of the last hidden layer, after a 0.3 dropout, and outputs a prediction for the binding affinity. We fine-tuned using a batch size of 12, an initial learning rate of 1\,$\times$\,10$^{-4}$ with linear decay, and AdamW optimization.

\subsection{Benchmarking of external predictors}
\label{sec:methods:benchmark}

MHCflurry 2.1.2 produces IC$_{50}$ values in nanomolar.  HLA-Apollo outputs raw logits that are proportional to binding likelihood, and these were used without further transformation.  MHCnuggets 2.3 already reports log$_{10}$ IC$_{50}$. HLA-Athena returns a score between 0 and 1, where larger values indicate stronger binders; we used the score as provided.  For NetMHCpan 4.2, we retained its percentile rank column; smaller ranks denote stronger predicted affinity and were incorporated directly into the analyses.  For every peptide–allele pair, predictions were paired with ground-truth log$_{10}$ IC$_{50}$ measurements (or with the qualitative labels described in Section~\ref{subsec:results:sensitivity}). 

\subsection{Code Availability}
Custom scripts and pipelines used for training and evaluation are publicly available at \url{https://github.com/sermare/ESMCBA}. 
\section{Acknowledgments}

This research used the Savio computational cluster resource provided by the Berkeley Research Computing program at the University of California, Berkeley (supported by the UC Berkeley Chancellor, Vice Chancellor for Research, and Chief Information Officer). NMI is a Chan Zuckerberg Biohub San Francisco Investigator.

\newpage
\bibliography{iclr2025_conference}

\clearpage
\appendix
\section*{Appendix: Supplementary Figures and Tables}

\renewcommand{\thetable}{S\arabic{table}}  % Renumber tables as S1, S2, ...
\setcounter{table}{0}  % Reset table counter

\begin{table}[b]        % add “bp” so LaTeX can place it at bottom if needed
  \centering
  \small
  % --- force centring even when the table is wider than \textwidth ---
  \makebox[\textwidth][c]{%
    \begin{tabular}{p{2.6cm} p{5.8cm} p{2.9cm} p{5.8cm}}
      \toprule
      \textbf{Model} & \textbf{Hidden / FC Layers} & \textbf{Total Parameters} &
      \textbf{Training Size / Dataset Size} \\ \midrule
      MHCnugget & 1 LSTM layer (64 units) and 1 fully connected layer (64 units) &
      $\sim$26 000 per allele-specific network & Varies by MHC allele; trained on IEDB 2018 data
      plus extra HLAp data for some alleles \\[3pt]

      HLApollo & 4 transformer encoder layers (400 dim, 16 heads) and 3 FC layers
      (256, 128, 1 units) & $\sim$11.7 million$^{*}$ &
      953 693 unique peptide–genotype tuples across 171 HLA-I alleles \\[3pt]

      MixMHCpred 2.2 & No hidden layers (position-weight matrices only) &
      $\sim$90 440$^{**}$ & 258 414 unique peptides, 384 070 peptide–HLA interactions,
      119 HLA-I alleles \\[3pt]

      HLAthena & 1 hidden layer (250 units, ReLU) & $\sim$4.3 million &
      186 464 unique peptides across 95 HLA-I alleles \\[3pt]

      MHCflurry 2.0 & 2–3 dense layers (256–1024 units, 50 \% dropout) &
      $\sim$355 841$^{***}$ & 713 069 peptide–MHC pairs across 171 HLA-I alleles \\[3pt]

      \textbf{ESMCBA} & 30 transformer encoder layers (960 dim, 20 heads) +
      linear prediction head & $\sim$333 million &
      Continued masked-language pre-training; supervised fine-tuning on peptide–MHC pairs
      across 121 HLA-I alleles \\[3pt]

      NetMHCpan 4.1 & Ensemble of 50 neural networks, each with 1 hidden layer
      (56 or 66 neurons) and 2 output neurons & $\sim$604 000 (estimated) &
      13 245 212 data points covering 250 distinct MHC class I molecules \\ \bottomrule
    \end{tabular}
  }

\caption{Model architectures, parameter counts, and training data for pMHC binding affinity predictors.  
$^{*}$ Estimate from HLApollo publication.  
$^{**}$ Sum of PWM parameters.  
$^{***}$ Reported in MHCflurry 2.0 release notes.}
\label{tab:model_comparison}
\end{table}

\clearpage
\begin{table}[p]
\centering
\caption{Predictive accuracy counts for recently submitted IEDB epitopes (2020 to 2025) across shared alleles}
\label{tab:recent_iedb_counts}
\begin{tabular}{lr}
\hline
\textbf{Allele} & \textbf{Peptides tested} \\
\hline
HLA-A*02:01 & \hphantom{0}400 \\
HLA-A*03:01 & \hphantom{0}139 \\
HLA-A*24:02 & \hphantom{0}120 \\
HLA-A*01:01 & \hphantom{0}115 \\
HLA-B*07:02 & \hphantom{0}101 \\
HLA-B*44:02 & \hphantom{00}97 \\
HLA-A*11:01 & \hphantom{00}92 \\
HLA-B*08:01 & \hphantom{00}43 \\
HLA-A*68:01 & \hphantom{00}30 \\
HLA-B*38:01 & \hphantom{00}12 \\
HLA-A*31:01 & \hphantom{000}8 \\
HLA-B*15:01 & \hphantom{000}8 \\
HLA-B*51:01 & \hphantom{000}7 \\
HLA-B*57:01 & \hphantom{000}7 \\
HLA-B*18:01 & \hphantom{000}7 \\
HLA-B*14:02 & \hphantom{000}6 \\
HLA-A*02:05 & \hphantom{000}3 \\
HLA-B*35:01 & \hphantom{000}2 \\
HLA-C*07:01 & \hphantom{000}2 \\
HLA-A*26:01 & \hphantom{000}2 \\
HLA-A*30:01 & \hphantom{000}2 \\
HLA-B*44:03 & \hphantom{000}1 \\
HLA-B*39:06 & \hphantom{000}1 \\
HLA-A*32:01 & \hphantom{000}1 \\
HLA-B*40:01 & \hphantom{000}1 \\
\hline
\end{tabular}
\end{table}
\clearpage

\begin{table}[p]
\centering
\caption{Number of peptides tested per allele for the ROC-AUC analysis of qualitative assay outcomes}
\label{tab:peptides_per_allele}
\begin{tabular}{lr}
\hline
\textbf{Allele} & \textbf{Peptides tested} \\
\hline
HLA-A*31:01 & 1000 \\
HLA-B*57:01 & 1000 \\
HLA-B*51:01 & 1000 \\
HLA-B*18:01 & 1000 \\
HLA-A*11:01 & 1000 \\
HLA-B*15:01 & 1000 \\
HLA-A*26:01 & 1000 \\
HLA-A*30:01 & 1000 \\
HLA-A*68:01 & 1000 \\
HLA-B*07:02 & 1000 \\
HLA-A*01:01 & 1000 \\
HLA-B*53:01 & 1000 \\
HLA-A*02:01 & 1000 \\
HLA-B*08:01 & 1000 \\
HLA-A*03:01 & 1000 \\
HLA-B*44:02 & 1000 \\
HLA-B*39:01 & 1000 \\
HLA-A*32:01 & \hphantom{0}838 \\
HLA-C*06:02 & \hphantom{0}221 \\
HLA-B*38:01 & \hphantom{0}165 \\
HLA-B*39:06 & \hphantom{00}45 \\
\hline
\end{tabular}
\end{table}

\end{document}